# Spatiotemporal focusing does not always improve axial intensity localization

Ya Cheng,[1,*] Hongqiang Xie,[1,2] Zhaohui Wang,[1,2] Guihua Li,[1] Bin Zeng,[1] Fei He,[1] Wei Chu,[1] Jinping Yao,[1] Lingling Qiao[1]

[1]*State Key Laboratory of High Field Laser Physics, Shanghai Institute of Optics and Fine Mechanics, Chinese Academy of Sciences, P.O. Box 800-211, Shanghai 201800, China*
[2]*Graduate School of the Chinese Academics of Sciences, Beijing 100039, China*
*\*Corresponding author: ya.cheng@siom.ac.cn*

We report on an experimental comparison on critical intensities of nonlinear self-focusing in air with conventional focusing and spatiotemporal focusing schemes. Our results show that the conventional focusing with the focal lens completely filled with the incident beam allows for the strongest axial intensity confinement against the self-focusing effect. This is because that in the high-numerical-aperture condition, the focal spot will have a compact size which results in a high focal intensity. Meanwhile, the Rayleigh length of the focused beam will be substantially shortened which efficiently postpones the onset of self-focusing.

Irradiation of tightly focused ultrafast laser pulses within transparent media results in strong axial confinement of the nonlinear laser-matter interaction near the focus, providing the enabling mechanism for three-dimensional (3D) micro- and nanostructuring [1] as well as 3D imaging of bio-tissues [2,3]. Typically, the tight focusing geometry requires the use of focal lenses of high numerical apertures (NAs), as these lenses are capable of producing tiny focal spots with sizes comparable to the laser wavelengths as well as short focal depths (i.e., high axial resolution). However, for high-throughput 3D microfabrication and bio-imaging of samples with large thickness, the high-NA lenses often suffer from their short working distances and small focal volumes. Under this circumstance, low-NA focal lenses, which offer higher penetration depths as well as larger sizes of the focal spots, seems to be a solution. Unfortunately, as the NA of focal lens decreases, the Rayleigh length, which is inversely proportional to the square of NA, will rapidly increase. More importantly, propagation of loosely focused intense laser pulses in transparent media can easily induce self-focusing effect, which can dramatically extend the focal depth in the samples [4,5]. The loss of axial intensity confinement under the low-NA condition has become a major challenge yet to be overcome.

Recently, several groups reported that strong suppression of the nonlinear self-focusing with a low-NA focal lens can be achieved using a spatiotemporal focusing scheme [6-12]. This effect has raised significant interest because of its potential in a wide variety of applications ranging from femtosecond laser micromachining, tissue ablation and laser-based remote sensing [13-17]. In the spatiotemporal focusing, the incident pulse is spatially chirped using a pair of gratings before entering the focal lens, which stretches the pulse width and substantially reduce the peak power of the laser beam. A dramatic temporal focusing (i.e., shortening of pulse duration) occurs around the focal spot because all the frequency components tend to recombine there, which restores the initial transform-limited pulses of the shortest duration at the focal point. We note that previously, the spatiotemporal focusing has been used in nonlinear multiphoton microscopy for performing optically sectioned wide-field bio-imaging, whereas the focal systems demonstrated there are unable to produce a small focal spots since the incident pulses are angularly but not spatially chirped [18-19].

Despite the experimental proofs on the enhancement of axial intensity confinement with the spatiotemporal focusing, doubts still remain. In fact, one may have noticed that almost all of the experiments mentioned above were performed in the condition of under filling of the lens aperture [6-12]. In the spatiotemporal focusing scheme, reduction of the incident beam diameter is unavoidable owing to the necessary spatial chirping of the incident pulse. However, for the conventional focusing scheme, it is well known that the strongest axial intensity confinement only occurs when the aperture of focal lens is completely filled. For linear propagation, Durfee et. al. have analytically shown that when a focal lens is completely filled, it always leads to a focal depth shorter than that of a spatiotemporally focused laser pulse regardless of the NA of the lens and the amount of spatial chirp of the incident pulse [8]. However, for applications such as micromachining and nonlinear optical imaging requiring high peak laser intensities at the focus, the situation becomes complicated as nonlinear effects will appear. In this Letter, we attempt to clarify whether in the nonlinear regime the spatiotemporal focusing can substantially improve the axial intensity confinement.

The dominating nonlinear effect affecting the axial intensity confinement is the self-focusing. Due to the spatial chirping, the peak power of the incident pulses in spatiotemporal focusing is much lower than that in conventional focusing. It is known that the self-focusing depends on the peak power but not the peak intensity of the pulses [4-5]. In addition, the spatial chirp of the incident pulses forms an elliptical incident beam shape, which further helps suppress the self-focusing [20-21]. The question is whether the gain in the axial intensity confinement contributed by the reduction of peak power and the change of beam profile from circular to elliptical with the spatiotemporal focusing could compensate for the loss in the axial intensity confinement due to the inevitable reduction of NA in the scheme. To find out the answer, we experimentally compare the focal intensities corresponding to the onset of nonlinear self-focusing of femtosecond laser pulses in air in the different focusing conditions. The comparison between the focal intensities but not the peak powers of the laser pulses will be more relevant to a variety of applications which require both high axial and lateral resolutions such as 3D nanofabrication and nonlinear optical imaging.

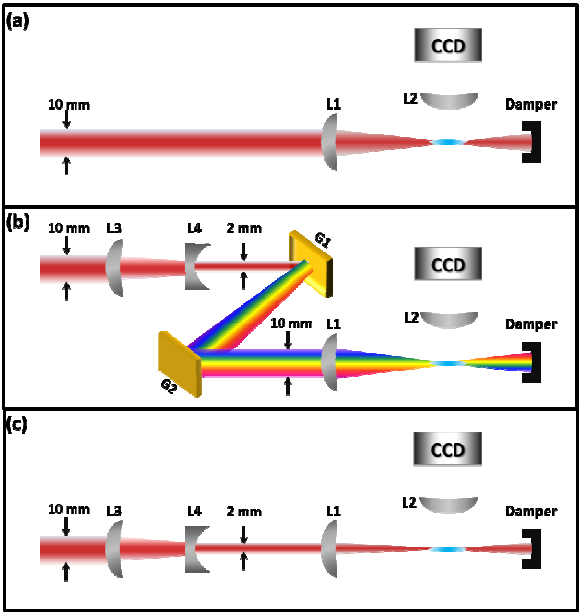

Fig. 1 Schematics of the Experimental setups. (a) Conventional focusing with a full-size beam. (b) Spatio-temporal focusing with a spatially chirped beam. (c) Conventional focusing with a beam of reduced diameter. BS: beam splitter, L1: focal lens (f = 4 cm), L2: imaging lens (f = 4 cm), L3 (f = 50 cm), L4 (f = -10 cm); G1, G2: Gratings (1500l/mm).

Figure 1 illustrates the experimental setup. The femtosecond laser system (Legend-Elite, Coherent Inc.) used in this experiment consists of a Ti:sapphire laser oscillator and amplifier, and a grating-based stretcher and compressor. In the conventional focusing scheme, transform-limited 800 nm, 50 fs, ~3 mJ pulses with a spectral bandwidth of ~26 nm at a 1-kHz repetition rate were directly focused in air to generate plasma, as illustrated in Fig. 1(a). The beam had a nearly Gaussian profile with a diameter of ~10 mm ($1/e^2$). In the spatiotemporal focusing, the diameter of the same amplified laser beam was first reduced to ~2 mm ($1/e^2$) using a telescope system consisting of a convex lens (f = 50 cm) and a concave lens (f = -10 cm). The laser pulses were then directed through a single-pass grating pair, which consisted of two σ = 1500 grooves/mm gratings blazing at 53°, to generate the spatial chirp (see Fig. 1(b)). More details on the arrangement of the gratings can be found elsewhere [22]. In the experiment, the spatially chirped beam had an elliptical beam profile with an ellipticity of 5. Thus, the major axis size of the spatially chirped pulses was the same as the beam diameter in the conventional focusing scheme. In addition, we also performed air plasma generation experiment by directly focusing the ~2 mm ($1/e^2$) laser beam into air, as shown in Fig. 1(c). In all the experiments, the pulse energy of femtoseocnd laser was controlled using a set of neutral density filters, and a focal lens with an aperture diameter of 25.4 mm and a focal length of 40 mm was used. In addition, the same focal lens was also used in combination with a charge coupled device (CCD) to capture the image of air plasma from the side of the optical axis, as illustrated in Fig. 1.

Figure 2 presents the side-view images of air plasma generated with the different focusing schemes. The laser pulse energies in the experiments were chosen based on the following consideration. First, we started from a very low pulse energy at which no air plasma could be observed. Then we gradually raised the pulse energy until a clear image of air plasma could be captured. From that point, we further increased the pulse energy with small steps to search for the signature of self-focusing. The onset of self-focusing was determined by observing a significant shift of focal spot toward the focal lens, as indicated by the guiding lines (the dashed curves) in Fig. 2. The pulse energy of each experiment was directly provided in the corresponding image of the generated plasma. It can be seen that in the conventional focusing with the full-size beam, the self-focusing started to occur at a pulse energy of ~60 μJ (Fig. 2(a)). In the spatiotemporal focusing, the self-focusing occurred when the pulse energy reached ~310 μJ (Fig. 2(b)). Lastly, for the conventionally focused beam with the reduced diameter of ~2 mm ($1/e^2$), onset of the self-focusing was observed at a pulse energy of ~75 μJ (Fig. 2(c)).

At first glance, the above results seemingly suggest that among all the focusing schemes illustrated in Fig. 1, the spatiotemporal focusing scheme can most efficiently prevent the self-focusing because of the highest critical power (i.e., pulse energy/pulse duration). However, for many nonlinear optical applications, it is the peak intensity rather than the peak power on target which plays the determining role. Since at the focal plane, both the conventionally and spatiotemporally focused beams have the same pulse duration, the peak intensities will be solely determined by the diameters of the focal spots. Assuming all the incident beams have an ideal Gaussian profile and the focal lens is aberration free, the focal spot diameters ($1/e^2$) in Fig. 1(a-c) are calculated to be ~4 μm, ~20 μm, and ~20 μm, respectively, in the linear propagation regime. Obviously, the conventional focusing scheme produces the smallest focal spot. Based on the pulse energies of self-focusing measured in Fig. 2(a-c), the corresponding peak intensities reach $1.9\times10^{16}$ W/cm$^2$, $0.4\times10^{16}$ W/cm$^2$, and $0.096\times10^{16}$ W/cm$^2$ for the conventional focusing with the full-size beam, the spatiotemporal focusing, and the conventional focusing with the beam of reduced diameter. It immediately becomes clear that the first focusing scheme can endure the highest peak intensity at focus against the nonlinear self-focusing, which is desirable by many practical applications. This is actually quite easy to understand because in the high-NA focusing condition, the focal spot has the smallest size, giving rise to high peak intensity; meanwhile, the Rayleigh length can be shortened which efficiently postpones the onset of self-focusing. One may argue that in reality, the high peak intensities calculated above by assuming an ideal focal spot can never be reached in air because of the strong plasma defocusing [23]. This is certainly true. However, the purpose of the current investigation is to provide a qualitative comparison of the potential capacities of axial intensity confinement between the conventional and spatiotemporal focusing schemes. The combined experimental measurements and the theoretical analysis have provided an unambiguous judgment on the dispute.

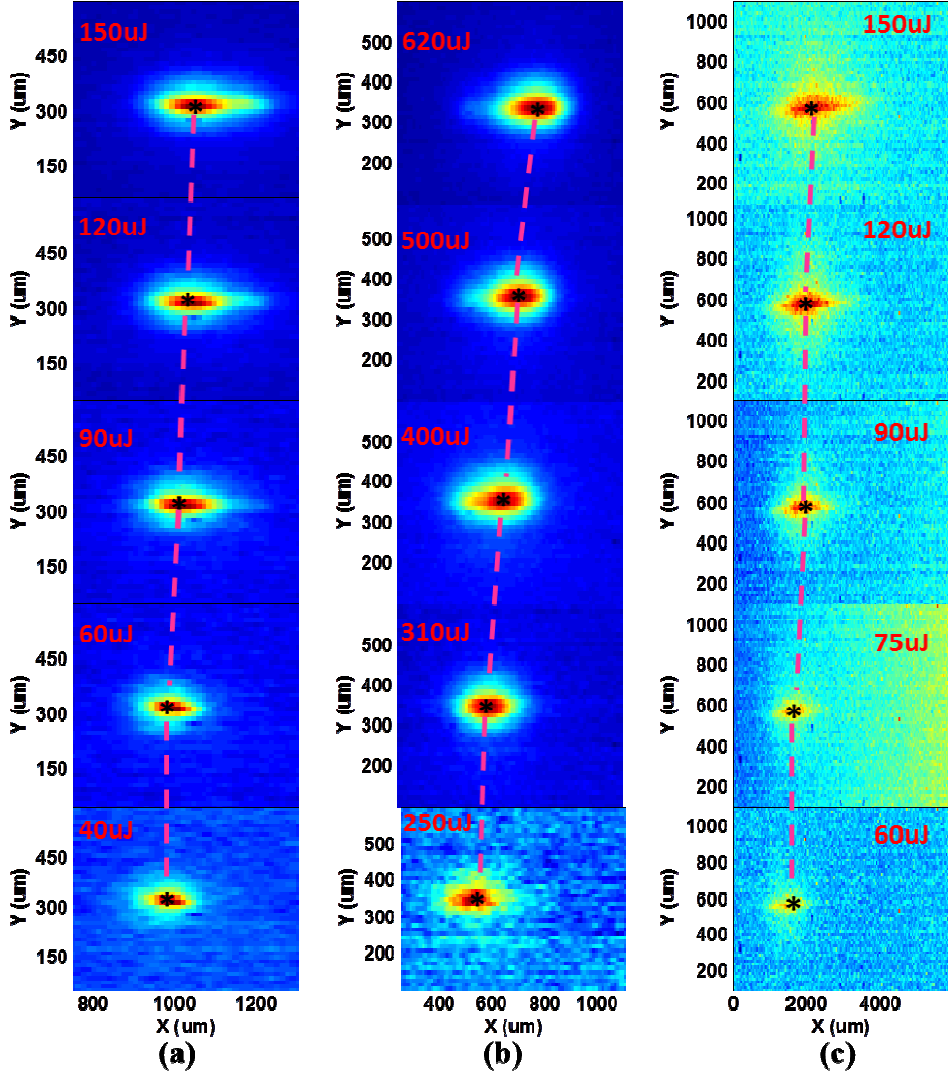

Fig. 2 Plasma luminescence captured with a CCD camera in different focusing conditions. (a) Conventional focusing with a full-size beam. (b) Spatio-temporal focusing with a spatially chirped beam. (c) Conventional focusing with a beam of reduced diameter. To facilitate the searching for the critical powers of self-focusing, the areas of the strongest luminescence are connected with dashed lines. Occurrence of self-focusing is determined by a sudden shift of the plasma toward the focal lens.

To conclude, we have investigated the nonlinear self-focusing of femtosecond laser pulses with the conventional and spatiotemporal focusing schemes. Our results show that for the applications pursuing high spatial resolutions in both lateral and axial directions, such as nonlinear optical imaging (e.g., SHG, THG, and CARS microscopy) and 3D nanofabrication (e.g., two-photon polymerization), the conventional focusing with the full-size beam outperforms the spatiotemporal focusing in terms of the ability to maintain the axial intensity localization against the nonlinear self-focusing. Although our experiments are carried out in air, the conclusion should hold for other transparent media. However, this does not mean that the spatiotemporal focusing is useless. Actually, as have been convincingly demonstrated before [13-17], once the strong axial intensity confinement should be achieved for a large focal spot, the spatiotemporal focusing becomes an ideal solution as it does not rely on tight focusing to reduce the focal depth. This unique characteristic opens the possibilities for high-throughput 3D materials processing, high-speed 3D bio-imaging, and high-sensitivity atmospheric remote sensing.

This work was supported by the National Basic Research Program of China (Grants No. 2014CB921300 and No. 2011CB808100), the National Natural Science Foundation of China (Grants No. 11134010 and No. 61275205).